\documentclass[12pt]{article}

\usepackage{amsmath}
\usepackage{amsfonts}
\usepackage{amssymb}
\usepackage{amsthm}
\usepackage{subfig}
\usepackage{float}
\usepackage{mathabx}
\usepackage[hidelinks]{hyperref}

\allowdisplaybreaks
\usepackage{natbib}
\usepackage{graphicx}

\usepackage[all]{nowidow}
\usepackage[margin=1.05in]{geometry}

\usepackage{tikz}
\usetikzlibrary{calc,fit,positioning}

\newtheorem{theorem}{Theorem}

\theoremstyle{remark}

\renewcommand{\hat}{\widehat}
\renewcommand{\bar}{\overline}

\DeclareMathOperator{\sign}{sign}

\DeclareMathOperator{\trace}{trace}
\renewcommand{\index}{\mathop{\mathrm{index}}}

\usepackage{setspace}
\title{A Note on Learnable Nash Equilibrium}  \author{Songzi Du\footnote{Du: Department
of Economics, University of California San Diego, sodu@ucsd.edu.}}
\date{June 21, 2026}

\begin{document}

\maketitle

%
%
%
%

\section{Introduction}

\citet*{Myerson1997} suggested the idea of sustainable equilibrium.  For example, in a coordination game

\begin{center}
\begin{tabular}{|c|c|c|}
  \hline
   & a & b \\ \hline
  a & (2,2) & (1,1) \\ \hline
  b & (1,1) & (2,2) \\
  \hline
\end{tabular}
\end{center}

The pure-strategy Nash equilibria (a,a) and (b,b) seem sustainable when the game is played in culturally familiar context, while the mixed-strategy Nash equilibirum (1/2 a + 1/2 b, 1/2 a + 1/2 b) does not seem sustainable.  One way to formalize sustainability, suggested by \citet*{Myerson1997}, is the notion of evolutionarily stable strategy \citep*{MaynardSmith1982}: a and b are evolutionarily stable strategies, while 1/2 a + 1/2 b is not.  Unfortunately an evolutionarily stable strategy often does not exist.  \citet*{Hofbauer2003SustainableLearnable} proposed a weakening of evolutionarily stable strategy: \emph{learnable Nash equilibrium}, which is asymptotically stable with respect to some myopic adjustment dynamic, a class of dynamics first studied by \citet*{Swinkels1993}; this is a weakening since an evolutionarily stable strategy is asymptotically stable under a variety of myopic adjustment dynamics.

\citet*{Hofbauer2003SustainableLearnable} stated a theorem that in a generic two-player game, a Nash equilibrium is learnable if and only if its index (as defined by \citet*{Shapley1974}, \citet*{Hofbauer_Sigmund_1998}) is +1.  However, the proof of the theorem has never been published and is not available in any format.  The only if direction of the theorem is implied by a general result of \citet*{Demichelis_Ritzberger_2003}.  In this note I give a simple proof of the if direction for symmetric two-player games using some basic linear algebra.  As is well known, for a generic game the Nash equilibria have either index +1 or -1, and the sum of indices is +1. Thus learnability is a powerful equilibrium refinement that exists in generic games and eliminates roughly half of the equilibria. This is in sharp contrast with the traditional refinement such as perfect, proper, and stable equilibria, which have no bite in generic normal-form games.  I hope this note would make the notion of learnable equilibrium more known and stimulate further developments.

\section{Model and result}\label{sec:model}

Consider a symmetric two-player (a.k.a.\ single population) game $A$: one's payoff is $A_{i,j}$ when one plays pure strategy $i$ and the opponent plays pure strategy $j$, where $1 \leq i, j \leq n$.

Let $\Delta = \{x \in \mathbb{R}_+^n : \sum_{i=1}^n x_i = 1\}$ be the set of mixed strategies (a.k.a.\ population states).  The mixed strategy $\hat{x} \in \Delta$ is a \emph{Nash equilibrium} if $\hat{x}' A\hat{x} \geq x' A \hat{x}$ for every $x \in \Delta$.  Here we take $\hat{x}$ and $x$ to be column vectors and $A$ a $n$-by-$n$ matrix.

Let $T$ be the tangent space to $\Delta$: $T = \{ x \in \mathbb{R}^n : \sum_{i=1}^n x_i = 0\}$.  We consider the adjustment dynamic given by $dx/dt = f(x)$, where $f: \Delta \to T$.  A \emph{myopic adjustment dynamic} is a mapping $f : \Delta \to T$ such that (i) $f$ is Lipschitz continuous, and $f_i(x) \geq 0$ if $x_i=0$, (ii) $f(x)' A x \geq 0$ for all $x \in \Delta$, and (iii) $f(x)=0$ if $x$ is a Nash equilibrium. Condition (i) ensures (by the Picard–Lindelöf theorem) that given any initial condition the differential equation $dx/dt = f(x)$ always has a unique solution $x(t)$, and the solution $x(t)$ stays in $\Delta$.  Condition (ii) says that the adjustment to $x$ is always in the direction of increasing the expected payoff: $(x+\epsilon f(x))' Ax \geq x' Ax$ for any $\epsilon > 0$; we call this the myopic adjustment condition.  Finally, condition (iii) says that a Nash equilibrium is a resting point of the dynamic $f$.

We say that a Nash equilibrium $\hat{x}$ is \emph{learnable} if there exists a myopic adjustment dynamic $f: \Delta \to T$ for which $\hat{x}$ is \emph{asymptotically stable}; that is, for every neighborhood $V$ of $\hat{x}$, there exists a neighborhood $U$ of $\hat{x}$ such that $x(0) \in U$ implies $x(t) \in V$ for all $t \geq 0$ and $\lim_{t \to \infty} x(t) = \hat{x}$, where $x(t)$ is the solution to
\[
\frac{dx(t)}{dt} = f(x(t)).
\]
Thus, learnability is a weak condition that just requires the equilibrium to be locally stable with respect to some myopic adjustment dynamic.  Nonetheless, as we would see, this weak condition would rule out roughly half of the equilibria in generic games.

An evolutionarily stable strategy $\hat{x}$ ($\hat{x}' A x > x' A x$ for every $x \neq \hat{x}$ in a neighborhood of $\hat{x}$) is a learnable Nash equilibrium: we can take $f$ to be the replicator dynamics ($f_i(x)=x_i((Ax)_i - x' Ax$)),\footnote{We can check that for the replicator dynamic, $f(x)'Ax = \sum_{i} (x_i ((Ax)_i)^2 - x_i (x'Ax) (Ax)_i) = \sum_{i} x_i ((Ax)_i)^2 - (x'Ax)^2 \geq 0$, so it is a myopic adjustment dynamic (condition (i) and (iii) are immediate).} among many other myopic adjustment dynamics; see \citet*{Hofbauer_Sigmund_1998}.  Moreover, \citet*{Demichelis_Ritzberger_2003} have proved that in a generic game if a Nash equilibrium is learnable, then it must have index +1.  In a generic symmetric two-player game $A$, if we normalize the payoffs so $A_{i,j} > 0$ for all $i$ and $j$, then the index of a Nash equilibrium $\hat{x}$ is
\[
\index(\hat{x}) = (-1)^{\#\hat{x}-1} \sign(\det A_{\hat{x}} ),
\]
where $\#\hat{x}$ is the size of the support of $\hat{x}$, and $A_{\hat{x}}$ is the submatrix of $A$ restricted to the support of $\hat{x}$.  This formula is originally obtained by \citet*{Shapley1974} for bimatrix games and specialized for symmetric games by \citet*{Balthasar2009}.  The genericity of $A$ implies that the index is either +1 or -1; a pure strategy Nash equilibrium has index +1, and the sum of the indices of all Nash equilibria is equal to +1 (\citet*{Balthasar2009}, Proposition 4.4).  For example, for the coordination game in the introduction, the a and b equilibria have index +1, while the 1/2 a + 1/2 b equilibrium has index -1.

In this paper we prove a converse to \citet*{Demichelis_Ritzberger_2003}:

\begin{theorem}
\label{thm:main}
For a generic $A$, if a Nash equilibrium has index +1, then it is learnable.
\end{theorem}

Consider a 3-by-3 coordination game:

\begin{center}
\begin{tabular}{|c|c|c|c|}
  \hline
   & a & b & c \\ \hline
  a & (2,2) & (1,1) & (1,1) \\ \hline
  b & (1,1) & (2,2) & (1,1) \\ \hline
  c & (1,1) & (1,1) & (2,2) \\
  \hline
\end{tabular}
\end{center}

It is easy to verify that the 1/3 a + 1/3 b + 1/3 c equilibrium has index +1.  However it is not learnable since this is a potential game, and any myopic adjustment dynamic moves in the direction of increasing the potential, which is maximized by the pure-strategy equilibria. Theorem \ref{thm:main} suggests that small payoff perturbations to this game would make the 1/3 a + 1/3 b + 1/3 c equilibrium learnable.

\section{A reformulation}
\label{sec:reformulation}

The tangent space $T$ is a $(n-1)$ subspace of $\mathbb{R}^n$.  So it is convenient to work with the first $(n-1)$ coordinates in $T$.

Fix a Nash equilibrium $\hat{x}$ with full support: $\hat{x}_i > 0$ for every $i =1, \ldots, n$.  Let $B$ be a $(n-1)$-by-$(n-1)$ matrix, where
\[
B_{i,j} = (A_{i,j} -A_{i,n}) - (A_{n,j} -A_{n,n}),
\]
for $1\leq i ,j \leq n-1$.

For $i = 1, \ldots, n-1$, let $u^i$ be the column vector where $u^i_j$ is $1$ if $j = i$, -1 if $j = n$, and 0 otherwise.  Let $U$ be the $n$-by-$n$ matrix whose $i$-th column is $u^i$, for $i = 1, \ldots, n-1$, and the $n$-th column is $\hat{x}$.  Then we have $\hat{B} = U' A U$, where
\[
\hat{B} = \begin{pmatrix}
B & 0 \\
0 & \hat{x}' A \hat{x}
\end{pmatrix}.
\]
If $A$ is normalized to have strictly positive payoffs, we have $\sign(\det B) = \sign(\det A)$.  Thus, $\index(\hat{x}) = (-1)^{n-1} \sign(\det B)$.

We can map $x \in \Delta$ to $y \in \mathbb{R}^{n-1}$, where $y_i = x_i - \hat{x}_i$ for $i =1 \ldots, n-1$, and vice versa when $y$ is sufficiently small. In words, $y$ is $x$'s displacement from $\hat{x}$.  Dynamics on $x$ ($dx/dt = f(x)$) and $y$ ($dy/dt = g(y)$) are related by
\begin{align}
\label{eq:gtof}
f_i(x) = g_i(y), \quad i = 1, \ldots, n-1; \quad f_n(x) = - \sum_{i=1}^{n-1} f_i(x).
\end{align}
We have
\begin{align}
\label{eq:adjustment_payoff}
f(x)' Ax = \sum_{i=1}^n f_i(x) ((A \hat{x})_i + (A (x-\hat{x}))_i) = g(y)' B y,
\end{align}
where we use the fact that $(A \hat{x})_i$ is constant over $i$.

Here is an intuitive argument for the necessity of index +1 for a learnable $\hat{x}$ (see \citet{Demichelis_Ritzberger_2003} for the rigorous proof). Suppose there exists a myopic adjustment dynamic $dx/dt = f(x)$ for which $\hat{x}$ is asymptotically stable (recall $\hat{x}$ is a fully mixed Nash equilibrium).  For the corresponding $dy/dt = g(y)$, $y=0$ is asymptotically stable.  Let us linearize the dynamic in a small neighborhood of $y=0$: $g(y) \approx D y$ for some non-singular $(n-1)$-by-$(n-1)$ matrix $D$.  Then the matrix $D$ is Hurwitz-stable: all eigenvalues of $D$ have negative real parts.  Thus, $\sign(\det D) = (-1)^{n-1}$.  Moreover, the myopic adjustment condition means that $(D y)' By \geq 0$ for all $y$ in the small neighborhood of 0.  This implies that $\det(D'B) = \det(D) \det(B) \geq 0$,\footnote{Suppose we have $y' C y \geq 0$ for every $y$ in a neighborhood of 0.  This implies that $\bar{C} = (C+C')/2$ is a positive semidefinite matrix since $C=\bar{C}+\tilde{C}$, where $\tilde{C} = (C-C')/2$, and $y' \tilde{C} y = 0$ for any $y$.  Suppose $\bar{C}$ is positive definite.  We write $C = \bar{C}^{1/2} (I + \bar{C}^{-1/2} \tilde{C} \bar{C}^{-1/2}) \bar{C}^{1/2}$, where $\bar{C}^{1/2}$ is symmetric, $\bar{C}^{1/2} \bar{C}^{1/2} = \bar{C}$ and $\bar{C}^{1/2}\bar{C}^{-1/2} = I$.  We have $\det(I + \bar{C}^{-1/2} \tilde{C} \bar{C}^{-1/2}) > 0$, since $\bar{C}^{-1/2} \tilde{C} \bar{C}^{-1/2}$ is a skew symmetric matrix so its eigenvalues are either 0 or purely imaginary.  Thus, $\det C > 0$.  If $\bar{C}$ is positive semidefinite but not positive definite, then we can use $\bar{C} = (C+C')/2 + \epsilon I$ for $\epsilon > 0$ and $\tilde{C} = (C-C')/2$ to conclude that $\det(C + \epsilon I) > 0$, and sending $\epsilon$ to zero implies that $\det C \geq 0$.} so $\sign(\det B) = \sign(\det D) = (-1)^{n-1}$, i.e., $\hat{x}$ has index +1.

\section{Proof of Theorem \ref{thm:main}}

For a generic $A$, every Nash equilibrium $\hat{x}$ is quasi-strict: $\hat{x}_i = 0 \Rightarrow (A\hat{x})_i < \hat{x}' A \hat{x}$ (\citet*{vonStengel2002}, Theorem 2.10).  So in a sufficiently small neighborhood of $\hat{x}$, if $x_i > 0$ but $\hat{x}_i=0$, we can set $f_i(x) < 0$ to drive $x_i$ to zero, while ensuring $f(x)' Ax > 0$, since $(Ax)_i < (A x)_j$ for $\hat{x}_j > 0$ if $x$ is sufficiently close to $\hat{x}$.  So without loss, we can assume that $\hat{x}$ has full support.  Suppose $\index(\hat{x}) = +1$, i.e., $\sign(\det(B)) = (-1)^{n-1}$, where $B$ is defined in Section \ref{sec:reformulation}.

We will construct
\[
g(y)= M B y,
\]
where $M$ is a $(n-1)$-by-$(n-1)$ matrix such that
\[
z' M z \geq 0,
\]
for every $z \in \mathbb{R}^{n-1}$, and $MB$ is Hurwitz-stable, i.e., all eigenvalues of $MB$ have negative real parts.

Using \eqref{eq:gtof}, \eqref{eq:adjustment_payoff}, and setting $z = By$, this gives a myopic adjustment dynamic $f(x)$ in a neighborhood of $\hat{x}$.  Since $MB$ is Hurwitz-stable, this implies $y=0$ is asymptotically stable for $dy/dt = g(y) = MB y$.  Thus, $x = \hat{x}$ is asymptotically stable for $dx/dt = f(x)$.

\subsection{The $n=3$ case}
\label{sec:proof:n3}
Let
\[
M = \alpha I + \beta J,
\]
where $\alpha>0$, $\beta \in \mathbb{R}$,
\[
I = \begin{pmatrix} 1 & 0 \\ 0 & 1 \end{pmatrix}, \quad J = \begin{pmatrix} 0 & -1 \\ 1 & 0 \end{pmatrix}.
\]

First, we check that
\[
z' M z = \alpha z' z \geq 0,
\]
since $z' J z = 0$ for every $z \in \mathbb{R}^2$.

A 2-by-2 matrix $MB$ is Hurwitz-stable if and only if $\det(MB) > 0$ and $\trace(MB) < 0$.
Now,
\[
\det((\alpha I + \beta J) B) = (\alpha^2 + \beta^2) \det B > 0,
\]
since $\det B > 0$ by assumption.

Suppose
\[
B = \begin{pmatrix}
      a & b \\
      c & d
    \end{pmatrix},
\]
then
\[
\trace((\alpha I + \beta J) B) = \alpha(a + d) +\beta(-c+b) < 0
\]
if $b \neq c$ and $\beta$ is sufficiently positive or negative.  And $b \neq c$ is clearly the generic case.

\subsection{The general case}

We now use the special case in Section \ref{sec:proof:n3} to prove the general case of Theorem \ref{thm:main}.  As before, let $\hat{x}$ be a Nash equilibrium with full support, with $(-1)^{n-1} \det(B) > 0$.

Using real Schur decomposition (\citet*{Horn_Johnson_2012}, Theorem 2.3.4), we can write $B = Q' T Q$, where $Q$ is a real orthogonal matrix ($Q'Q=I$), and $T$ is a real quasi-triangular matrix:
\[
T = \begin{bmatrix}
C_{11} & C_{12} & \dots & C_{1m} \\
0 & C_{22} & \dots & C_{2m} \\
\vdots & \vdots & \ddots & \vdots \\
0 & 0 & \dots & C_{mm}
\end{bmatrix},
\]
where each $C_{ii}$ is either a 1-by-1 matrix or 2-by-2 matrix.  In the former case, $C_{ii}$ is a real eigenvalue of $B$; in the latter case, $C_{ii}$ has non-real (conjugate) eigenvalues which are also eigenvalues of $B$.  We can rearrange the rows and columns to group the positive eigenvalues together, so that
\[
T = \begin{bmatrix}
T_{11} & T_{12} & \dots & T_{1k} \\
0 & T_{22} & \dots & T_{2k} \\
\vdots & \vdots & \ddots & \vdots \\
0 & 0 & \dots & T_{kk}
\end{bmatrix},
\]
where each $T_{ii}$ is either (i) a 2-by-2 upper triangular matrix whose diagonal are two positive eigenvalues of $B$, (ii) a 2-by-2 matrix with non-real (conjugate) eigenvalues which are also eigenvalues of $B$, or (iii) a 1-by-1 matrix which is a negative eigenvalues of $B$.  This can be done because $\sign(\det B) = (-1)^{n-1}$, so the number of positive eigenvalues of $B$ is always even.

If $T_{ii}$ is a 2-by-2 matrix, choose $K_i = (\alpha I + \beta J)$ as in Section \ref{sec:proof:n3} so that $K_i T_{ii}$ is Hurwitz-stable and $y' K_i y \geq 0$ for every $y \in \mathbb{R}^2$.  If $T_{ii}$ is a negative number, choose $K_i = 1$.  Let
\[
K = \begin{bmatrix}
K_{1} &0 & \dots & 0 \\
0 & K_{2} & \dots & 0 \\
\vdots & \vdots & \ddots & \vdots \\
0 & 0 & \dots & K_k
\end{bmatrix},
\]
we then have $y' K y = \sum_{i=1}^k y_i' K_i y_i \geq 0$ for every $y = (y_1, y_2, \ldots, y_k) \in \mathbb{R}^{n-1}$, and $K T$ is Hurwitz-stable, since $K T$ is a block upper triangular matrix with $K_i T_{ii}$'s on the diagonal, and a block upper triangular matrix is Hurwitz-stable if and only if each matrix on the diagonal is Hurwitz-stable\footnote{For a block upper triangular matrix e.g., $D = \begin{pmatrix}D_{11} & D_{12} \\ 0 & D_{22} \end{pmatrix}$, its characteristic polynomial is $\det(D - \lambda I) = \det(D_{11} - \lambda I) \det(D_{22}-\lambda I)$.  Hence all eigenvalues of $D$ have negative real parts if and only if the same is true for both $D_{11}$ and $D_{22}$.}.

Finally, we set
\[
M = Q' K Q.
\]
Clearly, we have $z' M z \geq 0$ for every $z \in \mathbb{R}^{n-1}$, and $MB$ is Hurwitz-stable.

\bibliography{references}

\end{document}